# Gyrotropic impact upon negatively refracting surfaces

**Allan Boardman[1,3], Neil King[1], Yuriy Rapoport[2] and Larry Velasco[1]**

[1]Photonics and Nonlinear Science Group, Joule Physics Laboratory, Institute for Materials Research, University of Salford, Salford, M54WT, UK

[2]National Taras Shevchenko University, Kiev, Ukraine

[3]Author to whom any correspondence should be addressed.

**Email: A.D.Boardman@salford.ac.uk, N.J.King@pgt.salford.ac.uk, l.n.velascohernandez@pgt.salford.ac.uk, laser@i.kiev.ua**

## Abstract

Surface wave propagation at the interface between different types of gyrotropic materials and an isotropic negatively refracting medium, in which the relative permittivity and relative permeability are, simultaneously, negative is investigated. A general approach is taken that embraces both the gyroelectric and gyromagnetic materials, permitting the possibility of operating in either the low GHz, THz or the optical frequency regimes. The classical transverse Voigt configuration is adopted and a complete analysis of nonreciprocal surface wave dispersion is presented. The impact of the surface polariton modes upon the reflection of both plane waves and beams is discussed in terms of resonances and an example of the influence upon the Goos-Hänchen shift is given.

## 1. Introduction

Surface waves have been a topic of research since the work of Lord Rayleigh on elastic solids [1]. They can exist under certain conditions on an entirely free surface bounded by air, or at the interface separating two semi-infinite half-spaces. Controlling the direction of such waves is very important for potential applications [2]. Normally the excitations are in the form of a guided wave bound to the surface, with the associated fields decaying exponentially along the normal directions, but they can assume the form of an unbound wave maintained by balancing the energy radiating away from the surface with incoming energy. Establishing a surface wave, therefore, depends on a critical number of issues and Rayleigh led the way in addressing them. From this pioneering period interest in surface waves grew in a number of directions and it has even been said about the subsequent era that surface waves sprouted like mushrooms [sic] [3], with as many as eleven different types being under discussion. However, the list was considerably inflated by including the geometry of the guiding surfaces. That does not mean that it is unimportant to discuss the surface sustaining the waves but it must be recognised that there is a generic underlying set of electromagnetic boundary conditions that determine the guided wave properties, and that these are common to all the geometries. By 1959 this had not been appreciated but by the 1970s a vigorous interest [4] developed in what came to be called surface polaritons. Although these excitations can indeed be sustained by using a variety of geometries, it is the planar geometry that has attracted most attention. To a certain extent this is because of the experimental ease with which the core properties can be exposed and exploited [5-8]. This is the geometry adopted here and the interfaces are between a gyrotropic material and a negative phase velocity metamaterial [9-20]. The early work on surface plasmon polaritons stimulated a lot of interest [21-25] in the role gyrotropy plays in modifying their fundamental properties. This has now led to the influence that an applied magnetic field has upon optical transmission through sub-wavelength hole arrays [26].

The first point to make about the basic theory of polaritons is that they arise when an electromagnetic wave passes through a polarizable dielectric and excites internal degrees of freedom. In an electron plasma a polariton mode has both a *photon* and a *plasmon* content, where a plasmon is the quantum of a coherent jelly-like oscillation of the whole electron sea. In their extreme states polaritons can have a high enough photon content to be treated as a light wave or a strong enough plasmon content to be just a plasma oscillation. A bulk plasmon mode has an angular frequency $\omega_p$ and a surface polariton in the same state has an oscillation angular frequency equal to $\omega_p / \sqrt{2}$. This was first shown by Ritchie [27] and one of the enduring features that has emerged from this field of investigation is that the methods of exciting the mixed modes called surface polaritons use attenuated, or frustrated, total reflection. This process is now globally referred to by the well known acronym **A**(ttenuated)**T**(otal)**R**(eflection). One form of this is called the Otto configuration [5], which involves using an isotropic prism under conditions of total internal reflection. The basis of ATR is that the total internal reflection total reflection is broken by the proximity of a metal surface [5], or the addition of a metal film to the base of a prism [6]. The latter configuration is known as the Kretchmann-Raether method. An analysis of these configurations can be carried out in a functional manner by calculating the reflectivity and showing that a minimum develops, beyond the critical angle, whenever a surface mode is developed. A more sophisticated argument identifies the generation of a surface polariton with the resonant behaviour of one of the Fresnel coefficient at the surface sustaining surface wave [4]. This resonance is associated with a strongly enhanced electric field at the boundary and is now used quite widely for detecting toxic chemicals and a wide variety of medical applications [2].

Since any controlling influence on the surface polaritons that are sustained by a negative phase velocity medium is going to be very important it is interesting to investigate the impact of an applied magnetic field upon their propagation characteristics. This is done through the agency of a gyrotropic material and, for this paper, the Voigt, or transverse field, configuration will be used. In this case, the applied magnetic field will be perpendicular to the propagation direction and lie in the plane defined by the interface. In the absence of a bounding negative phase velocity medium it is well known that a semi-infinite gyrotropic material will sustain non-reciprocal waves [21-25,28-35], so this is the major feature that the gyrotropic material brings to the TE and TM modes that are expected for the negative phase velocity medium [13]. The types of gyrotropic materials that can be used range from YIG, if low GHz operation [24] is required, through surface magneto-plasma waves [25] on semiconductors operating at THz frequencies, to magneto-optic layered systems designed to provide optical enhancement of Kerr phenomena [28]. The questions addressed throughout the paper concern the impact of the gyrotropic media on the permitted transparent/non-transparent surface polariton bands of the negative phase velocity medium, or, to turn the question around, the impact of the negative phase medium upon the non-reciprocal behaviour.

## 2. Theory of gyrotropic surface waves

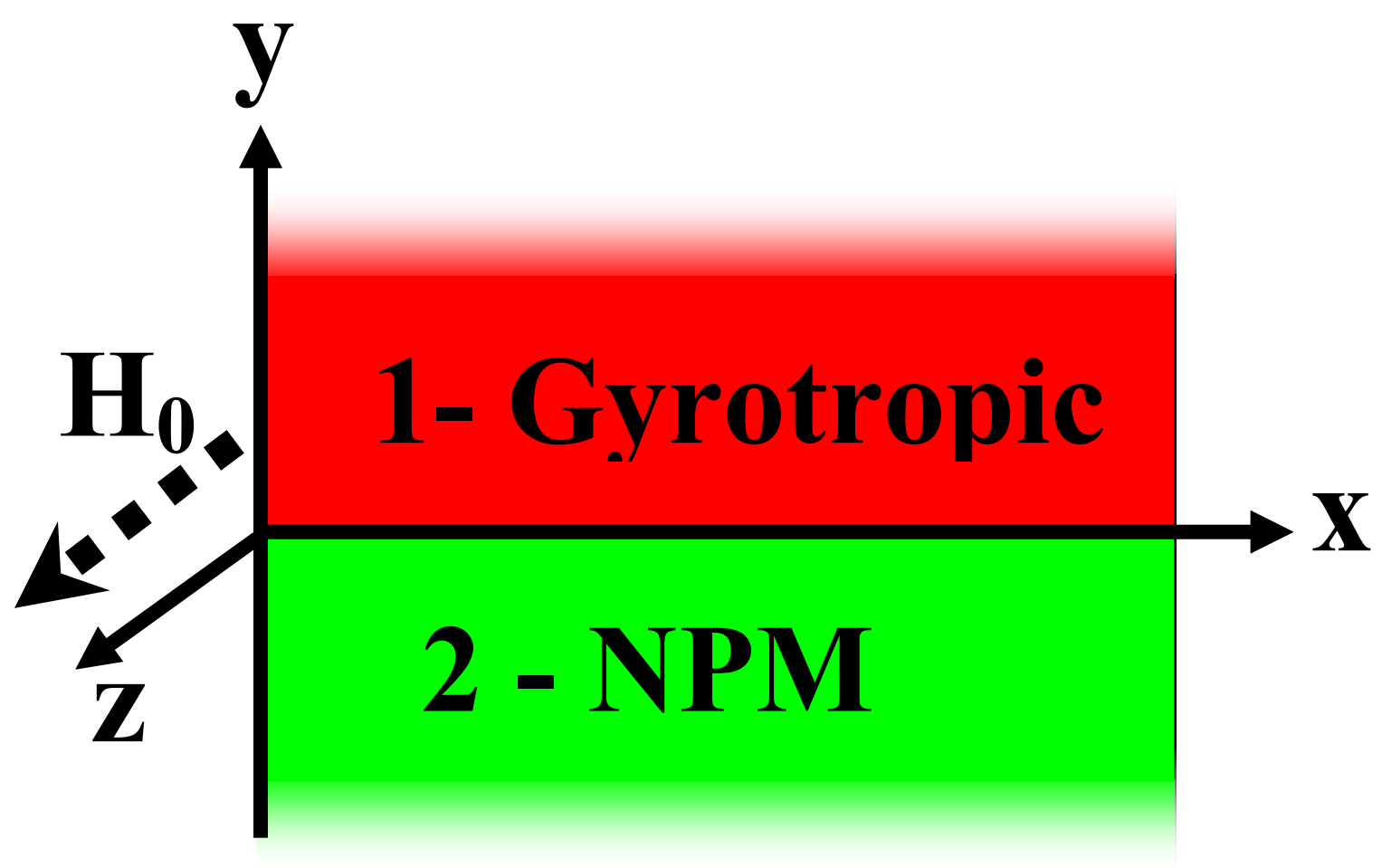

**Figure 1 Schematic diagram for an interface between a semi-infinite negative phase velocity medium [NPM] and a semi-infinite gyrotropic medium.**

An *interface* between semi-infinite gyrotropic and negative phase velocity (NPM) materials is sketched in figure 1. The latter also serves to define the coordinate system to be used in the rest of the paper and shows that a constant *applied* magnetic field $\mathbf{H}_0$ is directed along the *z*-axis. This is the so-called *transverse* or *Voigt* configuration. The surface waves propagate along the *x*-axis with a wave number $k_x$, and the attenuation of the waves in the *y*-directions are defined by the quantities $\kappa_{1,2}$. The electric and magnetic field vectors used here are

$$\left.\begin{aligned}\mathbf{E} &= \mathbf{E}_1 \exp\left[i\left(k_x x - \omega t\right) - \kappa_1 y\right] \\ \mathbf{H} &= \mathbf{H}_1 \exp\left[i\left(k_x x - \omega t\right) - \kappa_1 y\right]\end{aligned}\right\} \quad y > 0 \tag{1}$$

$$\left.\begin{aligned}\mathbf{E} &= \mathbf{E}_2 \exp\left[i\left(k_x x - \omega t\right) + \kappa_2 y\right] \\ \mathbf{H} &= \mathbf{H}_2 \exp\left[i\left(k_x x - \omega t\right) + \kappa_2 y\right]\end{aligned}\right\} \quad y < 0 \tag{2}$$

### *2.1. Gyroelectric waves*

Since this is the transverse, Voigt, case, the relative permittivity of medium 1 is a tensor, and for the coordinate system adopted is [32]

$$\boldsymbol{\varepsilon} = \begin{pmatrix} \varepsilon_{xx} & \varepsilon_{xy} & 0 \\ -\varepsilon_{xy} & \varepsilon_{xx} & 0 \\ 0 & 0 & \varepsilon_{zz} \end{pmatrix} \tag{3}$$

Note that the (*x*,*y*) diagonal elements are equal but unequal to the remaining element. The off-diagonal terms are a direct manifestation of the presence of the applied magnetic field. The electric field components in this gyroelectric medium depend upon each other in the following way

$$\begin{pmatrix} -\kappa_1^2 - \frac{\omega^2}{c^2}\mu_1\varepsilon_{xx} & ik_x\kappa_1 - \frac{\omega^2}{c^2}\mu_1\varepsilon_{xy} \\ ik_x\kappa_1 + \frac{\omega^2}{c^2}\mu_1\varepsilon_{xy} & k_x^2 - \frac{\omega^2}{c^2}\mu_1\varepsilon_{xx} \end{pmatrix}\begin{pmatrix} E_{x1} \\ E_{y1} \end{pmatrix} = 0 \tag{4}$$

$$\left[k_x^2 - \kappa_1^2 + \frac{\omega^2}{c^2}\mu_1\varepsilon_{zz}\right]E_{z1} = 0 \tag{5}$$

where $\mu_1$ has been included for generality, even though it is anticipated that medium 1 will have a relative permeability of unity, in practical cases. Equations (4) and (5) show that the TE mode $(E_z, H_x, H_y)$ is not coupled to the TM mode $(H_z, E_x, E_y)$ and that it is the *TM mode that is affected by the gyrotropy*. Furthermore, for the TM modes

$$\frac{\omega^2}{c^2}\mu_1\varepsilon_V - k_x^2 + \kappa_1^2 = 0 \tag{6}$$

in which the Voigt relative permittivity [33,34] is introduced as

$$\varepsilon_V = \varepsilon_{xx} + \frac{\varepsilon_{xy}^2}{\varepsilon_{xx}} \tag{7}$$

For the negative phase velocity medium, the attenuation function $\kappa_2$ is the solution of

$$\frac{\omega^2}{c^2}\mu_2\varepsilon_2 - k_x^2 + \kappa_2^2 = 0 \tag{8}$$

At the interface, the continuity of the *x*-components of the electric field vector $\mathbf{E}$, and the normal component of the displacement vector $\mathbf{D}$, lead to

$$\varepsilon_2 \frac{E_{y2}}{E_{x2}} = \varepsilon_{xx}\frac{E_{y1}}{E_{x1}} - \varepsilon_{xy} \tag{9}$$

In addition, from $\nabla \cdot \mathbf{D} = 0$,

$$\frac{E_{y2}}{E_{x2}} = \frac{ik_x}{\kappa_2} \tag{10}$$

$$\frac{E_{y1}}{E_{x1}} = \frac{\left(\kappa_1\varepsilon_{xy} - ik_x\varepsilon_{xx}\right)}{\left(\kappa_1\varepsilon_{xx} + ik_x\varepsilon_{xy}\right)} \tag{11}$$

Finally, the dispersion equation for the surface wave at the interface of semi-infinite gyroelectric and negative phase velocity media is

$$\frac{\kappa_2}{\varepsilon_2} + \frac{\kappa_1}{\varepsilon_V} + \frac{ik_x}{\varepsilon_V}\frac{\varepsilon_{xy}}{\varepsilon_{xx}} = 0 \tag{12}$$

*2.2. Gyromagnetic waves*

The dispersion equation for a gyromagnetic material can now be produced by replacing all the relative permittivities in (12) with the corresponding relative permeability tensor elements. The result is

$$\frac{\kappa_2}{\mu_2}+\frac{\kappa_1}{\mu_V}+\frac{ik_x}{\mu_V}\frac{\mu_{xy}}{\mu_{xx}}=0 \tag{13}$$

where [24]

$$\mu_{xx}=\frac{\omega_H(\omega_H+\omega_M)-\omega^2}{(\omega_H^2-\omega^2)},\ \mu_{xy}=i\frac{\omega_M\omega}{(\omega_H^2-\omega^2)},\ \mu_V=\frac{\mu_{xx}^2+\mu_{xy}^2}{\mu_{xx}} \tag{14}$$

in which $\omega_H=\gamma H_0$, $\omega_M=4\pi\gamma M_0$. $\omega_H$ is the natural precessional frequency of the magnetisation vector about a constant applied magnetic field $H_0$. $\omega_M$ is a frequency that depends upon the saturation magnetization $M_0$ of the material under investigation. The frequency of operation will be in the low GHz range, so the model adopted for the metamaterial is now [9-20]

$$\varepsilon_2=1-\frac{\omega_p^2}{\omega^2},\ \mu_2=1-\frac{F\omega^2}{\omega^2-\omega_0^2} \tag{15}$$

where F, $\omega_p$ and $\omega_0$ are disposable parameters. In the gyromagnetic case the surface polariton modes are TE-polarised, whereas waves on a gyroelectric material are TM-polarised. In fact the latter reduce to the familiar surface plasmon-polariton limit when the applied magnetic field is switched off. This is potentially very interesting because the negative phase velocity medium on its own supports *both* TE and TM-polarised surface polariton modes and region of transparency when the relative permittivity and relative permeability are *simultaneously* negative can be associated mainly with TE modes [13], or for different sets of data with the TM polarization. In fact the frequency bands are quite interchangeable. The mutual impact of gyrotropy and negative phase behaviour is therefore an important issue to address and can have a number of outcomes. Since reversing the magnetic field direction or propagation direction reverses the signs of $\varepsilon_{xy}$ and $\mu_{xy}$ the surface polariton modes on the combined material shown in figure 1 are also nonreciprocal.

*2.3. Gyrotropic surface wave dispersion: analytic form*

Starting with a gyroelectric material interface, a rearrangement of (12) gives

$$\left(k_x^2-k_0^2\mu_1\varepsilon_V\right)\varepsilon_2{}^2\varepsilon_{xx}{}^2=\left(k_x^2-k_0^2\mu_2\varepsilon_2\right)\varepsilon_V{}^2\varepsilon_{xx}{}^2+2ik_x\kappa_2\varepsilon_V\varepsilon_2\varepsilon_{xx}\varepsilon_{xy}-k_x^2\varepsilon_2{}^2\varepsilon_{xy}{}^2 \tag{16}$$

where $k_0=\omega/c$. The dispersion equation then becomes

$$k_x^2=k_0^2\frac{K_1\pm K_2}{\left[\left(\varepsilon_2{}^2-\varepsilon_V\varepsilon_{xx}\right)^2+4\varepsilon_2{}^2\varepsilon_{xy}{}^2\right]} \tag{17}$$

$$K_1 = \left[\left(\varepsilon_V\varepsilon_{xx} - \varepsilon_2{}^2\right)\left(\mu_2\varepsilon_V - \mu_1\varepsilon_2\right)\varepsilon_2\varepsilon_{xx} + 4\mu_2\varepsilon_2{}^3\varepsilon_{xy}{}^2\right]$$
$$K_2 = 2i\varepsilon_{xy}\varepsilon_2{}^2\left\{\varepsilon_2\varepsilon_{xx}\left[\varepsilon_2\varepsilon_{xx}\left(\mu_1{}^2 + \mu_2{}^2\right) - \mu_1\mu_2\left(\varepsilon_2{}^2 + \varepsilon_V\varepsilon_{xx}\right)\right]\right\}^{1/2} \tag{18}$$

The constraints that the $\kappa$ functions must be positive and real for the fields to decay exponentially into each medium means that the boundaries delineating the permitted regions of existence will be the lines along which the $\kappa$ functions are zero i.e. the bulk polaritons dispersion curves. The gyromagnetic case is obtained by performing the replacement $\varepsilon \leftrightarrow \mu$. In (17), however, *all* of the constitutive parameters have been included hence this result applies to any interface between two media. Equation (17) has not yet assumed a form for any of the constitutive parameters. In other words they could be constants or functions of frequency. They model a magneto-optic material, a magneto-plasma, a negative phase velocity medium, a magnetic medium, metals and dielectric materials.
A perturbation solution of (12) must involve $k_x^{(0)}$, which is the zero magnetic field value of the propagation wave number i.e.

$$k_x^{(0)2} = k_0^2\left(\frac{\varepsilon_2\varepsilon_{xx}[\varepsilon_{xx}\mu_2 - \varepsilon_2]}{[\varepsilon_{xx}^2 - \varepsilon_2^2]}\right) \tag{19}$$

If $\kappa_1^{(0)}$ is also introduced then the perturbed solution is simply

$$k_x = k_x^{(0)} + i\varepsilon_{xy}\left(\frac{\varepsilon_2^2}{\varepsilon_1^2 - \varepsilon_2^2}\right)\kappa_1^{(0)} \tag{20}$$

## 2.4. Group velocity

The group velocity of a surface polariton shows the nature of the cross-over between forward and backward waves. A transit between the forward and backward wave propagation directions will be influenced by important factors such as the relative dielectric permittivity value of the material that is lying upon the negative phase velocity medium. It has been established above that, for the Voigt effect, TM modes assume an important role while, for the *gyromagnetic* case it is the TE mode that is important. It is necessary, therefore, to investigate the criteria under which modes of either polarization can change sign. This will now be developed. First of all, the Drude model will be used but then a full generalisation will be given that will embrace the so-called F-model, often adopted at low GHz frequencies.

### Gyroelectric TE Modes

The dispersion equation is

$$D(\omega, k_x) = \frac{\kappa_1}{\mu_1} + \frac{\kappa_2}{\mu_2} = 0 \tag{21}$$

and the Drude model of the negative phase velocity medium is

$$\varepsilon_2 = 1 - \frac{\omega_{pe}^2}{\omega^2} \qquad \mu_2 = 1 - \frac{\omega_{pm}^2}{\omega^2} \tag{22}$$

provided that losses are ignored. From (21)

$$\frac{\partial D}{\partial \omega}\delta\omega+\frac{\partial D}{\partial k_x}\delta k_x=0 \quad \Rightarrow v_g=\frac{\partial \omega}{\partial k_x}=-\frac{\partial D/\partial k_x}{\partial D/\partial \omega} \tag{23}$$

For the TE modes, setting $\mu_1=1$,

$$\frac{\omega^2}{c^2}\varepsilon_{xx}-k_x^2+\kappa_1^2=0, \quad \frac{\omega^2}{c^2}\mu_2\varepsilon_2-k_x^2+\kappa_2^2=0 \tag{24}$$

After performing the differentiations

$$\frac{\partial D}{\partial k_x}=\frac{\partial D}{\partial \kappa_1}\frac{\partial \kappa_1}{\partial k_x}+\frac{\partial D}{\partial \kappa_2}\frac{\partial \kappa_2}{\partial k_x}, \qquad \frac{\partial D}{\partial \omega}=\frac{\partial D}{\partial \kappa_1}\frac{\partial \kappa_1}{\partial \omega}+\frac{\partial D}{\partial \kappa_2}\frac{\partial \kappa_2}{\partial \omega} \tag{25}$$

and then completing some further manipulations, the expression for the group velocity becomes

$$v_g=\frac{k_x\left(\dfrac{1}{\mu_1\kappa_1}+\dfrac{1}{\mu_2\kappa_2}\right)}{\dfrac{\omega}{c^2}\left(\dfrac{\varepsilon_{xx}}{\kappa_1}+\dfrac{1}{\mu_2\kappa_2}\right)-\dfrac{\omega_{pm}^2}{\omega^3\mu_2\kappa_2 c^2}[\omega_{pe}^2-2\dfrac{\kappa_2 c^2}{\mu_2}]} \tag{26}$$

in which $\mu_1=1$ for a magneto-optic or magneto-plasma material. The group velocity in this case does not depend upon the gyroelectric properties of the medium bounding the negative phase medium but it is interesting that that it goes to zero and changes sign whenever the bounding dielectric medium of the negative phase velocity material has a relative permittivity $|\varepsilon_1|=|\varepsilon_2||\mu_2|$.

## TM Modes

In this case, the gyroelectric properties will be become apparent through the dispersion equation

$$D(\omega,k_x)=\frac{\kappa_1\varepsilon_{xx}+i\varepsilon_{xy}k_x}{\varepsilon_{xx}^2+\varepsilon_{xy}^2}+\frac{\kappa_2}{\varepsilon_2} \tag{27}$$

Following a similar development to the TE case, the group velocity is

$$v_g=\frac{k_x\left(\dfrac{\varepsilon_{eff}}{\kappa_1}+\dfrac{1}{\varepsilon_2\kappa_2}\right)+\dfrac{i\varepsilon_{xy}}{\varepsilon_{xx}^2+\varepsilon_{xy}^2}}{\dfrac{\omega}{c^2}\left(\dfrac{\varepsilon_{eff}^2}{\kappa_1}+\dfrac{1}{\mu_2\kappa_2}\right)-\dfrac{\omega_{pe}^2}{\omega^3\varepsilon_2\kappa_2 c^2}[\omega_{pm}^2-2\dfrac{\kappa_2^2 c^2}{\varepsilon_2}]} \tag{28}$$

The foregoing special cases can be drawn out of a general formula, which is

$$v_g = \frac{k_x\left(\frac{1}{\mu_2 k_{y2}}+\frac{1}{\mu_V k_{y1}}\right)+\frac{i}{\mu_V}\frac{\mu_{xy}}{\mu_{xx}}}{\frac{k_{y2}}{\mu_2^2}\frac{\partial\mu_2}{\partial\omega}-\frac{1}{\mu_2}\frac{\partial k_{y2}}{\partial\omega}+\frac{k_{y1}}{\mu_V^2}\frac{\partial\mu_V}{\partial\omega}-\frac{1}{\mu_V}\frac{\partial k_{y1}}{\partial\omega}+ik_x\left(\frac{\left(\mu_{xx}{}^2-\mu_{xy}{}^2\right)\frac{\partial\mu_{xy}}{\partial\omega}-2\mu_{xx}\mu_{xy}\frac{\partial\mu_{xx}}{\partial\omega}}{\left(\mu_{xx}{}^2+\mu_{xy}{}^2\right)^2}\right)} \quad (29)$$

This equation, allows both the low GHz frequency range, driven by the F-model, and the high frequency ranges more appropriate to magnet-optic and magneto-plasma bounding media to be developed.

# 3. Reflection of plane waves and beams

The use of gyrotropic materials increases the number of applications [28] of a given integrated optics device by adding design capability that comes from the application of an applied magnetic field. Suppose that a plane TM-polarised wave is travelling in glass towards a single planar loss-free glass-air interface, separating semi-infinite regions. It will experience a reflectivity that drops to zero at the Brewster angle and then rises until the critical angle is reached. It then remains at unity. For an air-metal interface, the relative dielectric function of the metal is $\varepsilon < 0$, where $\varepsilon$ is a function of frequency and is complex due to absorption. In this case, a plane wave incident through glass onto the metal will experience a dip in the reflectivity at what can be called a pseudo-Brewster angle, before then rising towards unity. If a prism made of glass, or some other material, is still used to carry an incident plane wave but now a semi-infinite metal is brought into proximity with the base of the prism that is reflecting the wave, or a thin film of metal is deposited onto the base of the prism, then localised waves can be launched onto the free surface of the metal. The signature of this launch is a sharp drop in the reflectivity just beyond the critical angle. Beyond this resonance angle, the reflectivity will rise again as described earlier. For a metal described by the complex relative dielectric function $\varepsilon = \varepsilon' + i\varepsilon''$, the half-width of the reflectivity minimum goes to zero if the damping is very small, so without the use of a *complex* relative permittivity the resonance will not be seen. Many of these arguments will apply to the use of negative phase media and it will always be the case that using small damping coefficients will produce sharply defined resonances but that damping can lead to large line widths and hence reduce the *quality* of the resonance. Since this paper sets out to address basic questions about the impact of gyrotropic materials in combination with negative phase velocity media, attention will be focused upon the classical ATR configuration consisting of an isotropic dielectric prism-gyrotropic-NPM prism arrangement. Even if this does not produce the classic production of localised surface waves and instead produces evidence of a pseudo-Brewster effect [19] it is the first step towards optimising the role of the gyromagnetic material. Recent further steps towards such optimisation [28] involve a complicated multi-layer arrangement but this will be the topic of a future publication.

### *3.1. Coefficient of reflection for plane waves*

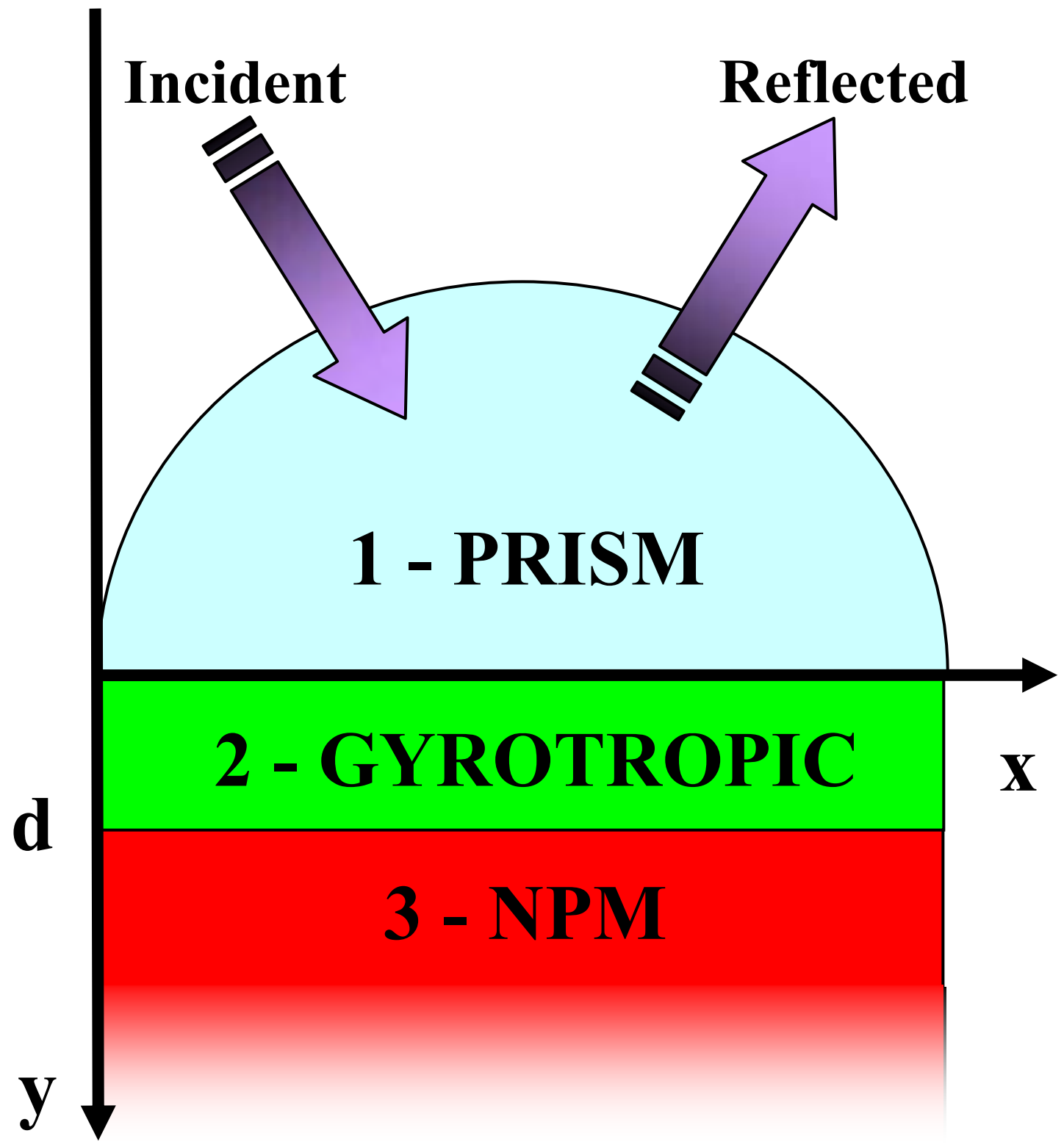

**Figure 2 Schematic diagram for an ATR configuration to generate surface waves at the interface between a negative phase velocity medium and a gyrotropic medium.**

Figure 2 shows a typical ATR arrangement [5,6] which consists of a prism-gyrotropic material-negative phase velocity material system, with the implication that total internal reflection within the prism will be frustrated by the proximity of a gyrotropic - NPM combination. If resonances do occur then they will be the signature of surface waves being generated at the gyrotropic - NPM interface. The system is invariant with respect to *z* and the magnetic field components along the *z*-axis are

$$H_{z1} = \left(A_{1+}\exp\left[ik_{y1}y\right] + A_{1-}\exp\left[-ik_{y1}y\right]\right)\exp\left[i\left(k_x x - \omega t\right)\right] \quad (30)$$

$$H_{z2} = \left(A_{2+}\exp\left[ik_{y2}y\right] + A_{2-}\exp\left[-ik_{y2}y\right]\right)\exp\left[i\left(k_x x - \omega t\right)\right] \quad (31)$$

$$H_{z3} = A_3\exp\left[ik_{y3}y\right]\exp\left[i\left(k_x x - \omega t\right)\right] \quad (32)$$

The tangential electric field components are

$$E_{x2} = [-\frac{q_+}{\omega\varepsilon_0}A_{2+}\exp\left[ik_{y2}y\right] + \frac{q_-}{\omega\varepsilon_0}A_{2-}\exp\left[-ik_{y2}y\right]]\exp\left[i\left(k_x x - \omega t\right)\right] \quad (33)$$

with the definitions

$$q_+ = \frac{\varepsilon_{xx}k_{y2} + k_x\varepsilon_{xy}}{\varepsilon_{xx}{}^2 + \varepsilon_{xy}{}^2}, \quad q_- = \frac{\varepsilon_{xx}k_{y2} - k_x\varepsilon_{xy}}{\varepsilon_{xx}{}^2 + \varepsilon_{xy}{}^2} \quad (34)$$

$$E_{x1} = -\frac{k_{y1}}{\omega\varepsilon_0\varepsilon_1}\left(A_{1+}\exp\left[ik_{y1}y\right] - A_{1-}\exp\left[-ik_{y1}y\right]\right)\exp\left[i\left(k_x x - \omega t\right)\right] \quad (35)$$

$$E_{x3} = -\frac{k_{y3}}{\omega\varepsilon_0\varepsilon_3} A_3 \exp\left[ik_{y3}y\right]\exp\left[i\left(k_x x - \omega t\right)\right] \quad (36)$$

The requirement that these tangential field components be continuous at the interfaces at $y$ = 0 and $d$ leads to the complex reflection coefficient $R_{TM} = \frac{A_{1-}}{A_{1+}}$, where

$$R_{TM} = \frac{A_{1-}}{A_{1+}} = \frac{\left[\frac{k_{y1}}{\varepsilon_1} - q_+\right]\left[\frac{k_{y3}}{\varepsilon_3} + q_-\right]\exp\left[-ik_{y2}d\right] - \left[\frac{k_{y1}}{\varepsilon_1} + q_-\right]\left[\frac{k_{y3}}{\varepsilon_3} - q_+\right]\exp\left[ik_{y2}d\right]}{\left[\frac{k_{y1}}{\varepsilon_1} + q_+\right]\left[\frac{k_{y3}}{\varepsilon_3} + q_-\right]\exp\left[-ik_{y2}d\right] + \left[q_- - \frac{k_{y1}}{\varepsilon_1}\right]\left[\frac{k_{y3}}{\varepsilon_3} - q_+\right]\exp\left[ik_{y2}d\right]} \quad (36)$$

This is the ATR reflection coefficient for a prism–gyrotropic medium–NPM configuration in the presence of a transversely applied external magnetic field.
The reflectivity from a prism–NPM–gyrotropic configuration can be obtained in a straightforward way based upon the analysis given above. The result is

$$R_{TM} = \frac{A_{1-}}{A_{1+}} = \frac{\left(\frac{k_{y1}}{\varepsilon_1} - \frac{k_{y2}}{\varepsilon_2}\right)\left(\frac{k_{y2}}{\varepsilon_2} + p\right)\exp\left[-ik_{y2}d\right] + \left(\frac{k_{y1}}{\varepsilon_1} + \frac{k_{y2}}{\varepsilon_2}\right)\left(\frac{k_{y2}}{\varepsilon_2} - p\right)\exp\left[ik_{y2}d\right]}{\left(\frac{k_{y1}}{\varepsilon_1} + \frac{k_{y2}}{\varepsilon_2}\right)\left(\frac{k_{y2}}{\varepsilon_2} + p\right)\exp\left[-ik_{y2}d\right] + \left(\frac{k_{y1}}{\varepsilon_1} - \frac{k_{y2}}{\varepsilon_2}\right)\left(\frac{k_{y2}}{\varepsilon_2} - p\right)\exp\left[ik_{y2}d\right]} \quad (37)$$

$$p = \frac{\left(\varepsilon_{xy}k_x + \varepsilon_{yy}k_{y3}\right)}{\left(\varepsilon_{xx}\varepsilon_{yy} - \varepsilon_{xy}\varepsilon_{yx}\right)} \quad (38)$$

### *3.2. The Goos-Hänchen shift of a beam*

If an electromagnetic wave that is travelling in a dense transparent medium, such as glass, encounters an interface to a less dense medium, like air, then it will be totally internally reflected if the angle of incidence is greater than the critical angle. An elementary plane wave is reflected as a plane wave and the electromagnetic field in the less dense medium decays exponentially in a direction normal to the interface. This logic works well until a finite beam, such a laser beam, is incident upon the interface. By examining the Fourier transform of a beam with a typical Gaussian profile, for example, it is easily appreciated that it is made up of an infinite number of plane waves: all with different amplitudes and directions. This means that upon observing the behaviour of real finite width beams as they reflect from a surfaces is different form that of a plane wave. In fact, they are laterally shifted from the commonly expected geometric optics position found for a plane wave. This *lateral* shift [36-38] is now eponymously known as the Goos-Hänchen shift [36], even though Newton had already discussed it [39].

In Figure 3a beam of electromagnetic radiation is falling upon a plane surface located at a point on the y-axis, some unidentified distance below the plane $y = 0$. Upon encountering the interface, total internal reflection occurs. The angle of incidence that the principal axis of the beam assumes to the normal is $\theta$. As shown in the sketch, the beam is also assumed, without loss of generality, to be Gaussian in shape. If a plane wave is used, however, instead of a beam, then the incident medium would be characterized by a wave number $k$: so $k$ serves to define the linear refractive index of the incident medium as $n = (\omega/c)k$, where ω is the angular frequency and c is the velocity of light in a vacuum [38]. If the incident medium is more dense than the medium below the interface then total internal reflection is expected on geometrical grounds.

The behaviour of the incoming and reflected beams can be analysed, to a large extent, analytically if they are assumed to have a Gaussian profile, which is one of the drivers for the choice of beam shape. The derivation presented here, is simplified by concentrating upon the essential features rather than the detailed manipulations that are in any case very straightforward [38]. First of all, local coordinate systems ($x_i$, $y_i$) and ($x_r$, $y_r$) with origins at the centre of the incident and reflected beams, respectively, ware adopted. By letting the plane y = 0 pass through the origin of the beam, rather than letting this plane be some notional height above an interface, as was done in a previous derivation [34] a considerable simplification ensues. Given that the plane $y = 0$ does, indeed, pass through the origin of the beam then the laboratory coordinates ($x$, $y$) are related to the local coordinates ($x_i$, $y_i$) for the incoming beam by the equations,

$$x_i = x\cos\theta, \quad y_i = x\sin\theta \tag{39}$$

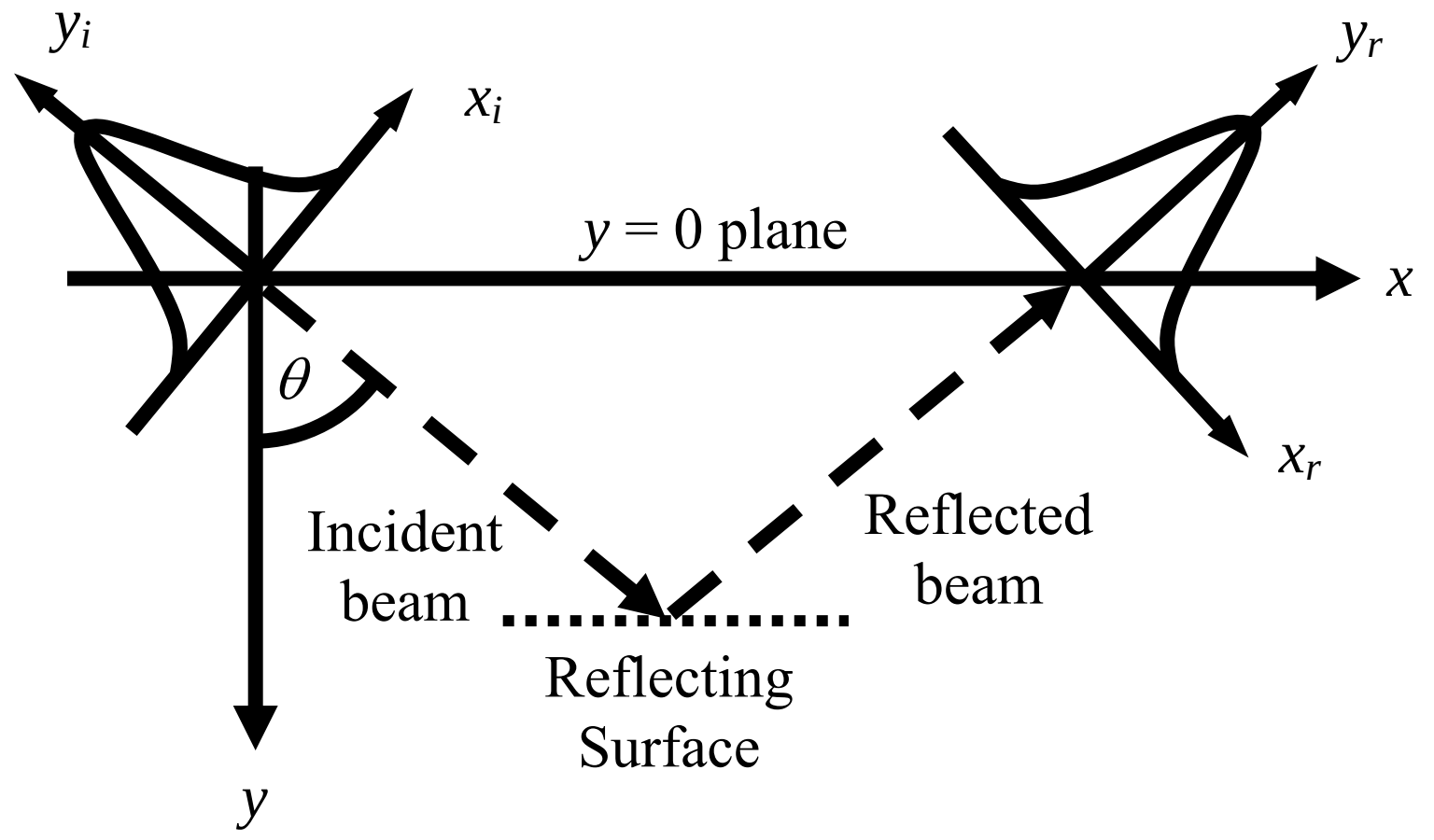

**Figure 3 Sketch of a Gaussian shaped beam totally reflecting from a surface**

If the beam is TM polarised then it is useful to focus the argument upon the remaining *magnetic field* component, H, which is perpendicular to the incident plane. This is assumed, in the local coordinate system, to have the Gaussian shape

$$H\,(x_i, y_i) = \frac{1}{\sqrt{\pi}w}\exp\left\{-\left(\frac{x_i}{w}\right)^2\right\} \tag{40}$$

This is a standard normalized form that has a $\sqrt{\pi}w$ factor introduced simply to make the integration over the range $-\infty \le x_i \le \infty$ equal to unity. $w$ is the half-width of the beam . In this local coordinate system, the beam propagates along the $y_i$ axis and it is a reasonable assumption that the beam is sufficiently compact in space for the fast variation to be the plane wave $\exp\{iky_i\}$, all the way across the Gaussian profile. Hence the projection onto the plane at $y = 0$, which is parallel to the physical interface, is

$$H(x,y=0)=\frac{1}{\sqrt{\pi}w}\exp\left\{-\left(\frac{x\cos\theta}{w}\right)^2\right\}\exp\{ikx\sin\theta\} \tag{41}$$

Now a beam can be thought of as an infinite set of plane waves: all travelling in different directions, with wave numbers ($k_x$, $k_y$). In other words, the Fourier transform of the incoming beam is

$$H_{inc}(x,y)=\frac{1}{2\pi}\int_{-\infty}^{+\infty}\Phi(k_x)\exp\{i[k_x x+k_y y]\}dk_x \tag{42}$$

since $k_y$ is a function of $k_x$, through the relationship $k_x{}^2+k_z{}^2=k^2$ and the Fourier amplitudes, after a straightforward integration, are

$$\Phi(k_x)=\int_{-\infty}^{+\infty}H(x,y=0)\exp\{-ik_x x\}dx=\frac{\exp(-\{[(k_x-k\sin\theta)/2\cos\theta\}^2}{\cos\theta} \tag{43}$$

The integral is at standard one, easily obtainable from standard tables. The *reflected* beam is

$$H_{refl}(x,z)=\frac{1}{2\pi\cos\theta}\int_{-\infty}^{+\infty}\Gamma(k_x)\exp\left\{-\left[\frac{(k_x-k\sin\theta)w}{\cos\theta}\right]^2\right\}\exp\{i[k_x x-k_z z]\}dk_x \tag{44}$$

where $\Gamma(k_x)=R_{TM}$ is the *plane wave* reflectivity, derived earlier and it should be emphasized that $\theta$ is the geometric angle of incidence i.e. the angle of incidence assumed by a plane wave travelling in the direction of the beam principal axis.

The next step is to choose an angle of incidence $\theta_0$, for which the tangential component of the wave number is $k_{x0}=k\sin\theta_0$, and then investigate the immediate vicinity of $\theta_0$ by means of a Taylor expansion. This means that $\Delta k_x=k_x-k\sin\theta_0$ in equation (44) and that the reflected amplitude is approximated by [40]

$$\Gamma(k_x)=\Gamma(k_{x0})+\left(\frac{\partial\Gamma(k_{x0})}{\partial k_x}\right)_{k_x=k_{x0}}\Delta k_x+..... \tag{45}$$

while the $y$-component of the wave number is approximated up to the second order i.e.

$$k_z\approx k_{z0}+\left(\frac{\partial k_z}{\partial k_x}\right)_{k_{z0}}\Delta k_x+\frac{1}{2}\left(\frac{\partial^2 k_z}{\partial k_x^2}\right)_{k_{z0}}(\Delta k_x)^2+..=k\cos\theta_0-\tan(\theta_0)(\Delta k_x)-\frac{1}{(2k\cos^3(\theta_0))}(\Delta k_x)^2 \tag{46}$$

Using these approximations in (43), together with some standard integrations and concentrating upon the reflected beam with a local coordinates system ($x_r$, $y_r$) leads directly to the result

$$H_{refl}(x_r) = H_{ro}\exp(-[x_r - i\Delta/2]^2/w^2)\exp(-\Delta^2/4w^2) \tag{47}$$

This shows that the Goos-Hänchen shift parallel to the $x_r$-axis is the real part of $i\Delta/2$. After evaluating all the integrals (47) shows that the shift $G$ is proportional to a combination of $\Gamma(k_x)$ and its derivative i.e.

$$G \propto S_{GH} = \mathrm{Re}[i\frac{1}{\cos\theta_0\Gamma(\theta_0)}\left(\frac{\partial\Gamma}{\partial\theta}\right)_{\theta_0}] = \frac{1}{\cos\theta_0}\mathrm{Re}[i\frac{1}{R_{TM}(\theta_0)}\left(\frac{\partial R_{TM}}{\partial\theta}\right)_{\theta_0}] \tag{48}$$

In the early days, the Goos-Hänchen was investigated near to the critical angle [38]. For example, in a glass prism undergoing total internal reflection this is really the only interesting angle. If there is the possibility of generating surface modes then this will occur well beyond the critical angle i.e. in the neighbourhood of a resonance, where the reflectivity plunges towards zero and the gradient is very steep. It is expected therefore that the shift associated with this will be very large [40]. This will be true whether the resonance is in a negative phase medium or not: so the "giantness" of the shift [41] is associated, almost entirely, with the resonance and not just the material nature.

$$S_{GH} = \frac{1}{\cos\theta_0\{(R_{TM}^r)^2 + (R_{TM}^i)^2\}}[R_{TM}^i\frac{\partial R_{TM}^r}{\partial\theta} - R_{TM}^r\frac{\partial R_{TM}^i}{\partial\theta}] \tag{49}$$

# 4. Numerical results

## 4.1. Semi-infinite Negative Phase Medium bounded by YIG

The dispersion equation that yields the properties of surface waves at the interface between semi-infinite media consisting of a negative phase velocity medium (NPM) and an insulating magnetic material called yttrium iron garnet, usually referred to as YIG, is given by equation (13). Since YIG operates effectively in the low GHz microwave frequency range, the model adopted here will the one defined by equation (14). At such frequencies the NPM metamaterial is modelled by (15). The parameters for the latter model can be selected in any way that is desired but the outcome is always that the NPM alone supports *both* TE- and TM-polarised surface waves. This is the signature of this type of metamaterial because, if $\mu_2 = 1$, only TM surface plasmon-polaritons are allowed. A very early data selection [13] produced permissible bands of frequencies that consisted of *forward* TM waves and *backward* TE waves and the TE polarised waves lay completely within the transparent negatively refracting region. The lower TM band lay completely within a pseudo-metallic region and the upper TM band only lay partially within the transparent negatively refracting region. Since these features depend upon the choice of data, there is no reason why the TE and TM bands for surface waves travelling on such a metamaterial cannot be interchanged. Furthermore, both the TE and TM branches can be brought entirely within the transparent region.

The dispersion equation can be expanded into the biquadratic equation

$$Ak_x^4 + Bk_x^2 + C = 0 \tag{50}$$

Since the solution is

$$k_x^2 = \frac{-B \pm \sqrt{D}}{2A}, D = B^2 - 4AC \tag{51}$$

where

$$A = \left(\mu_2{}^2 - \mu_V \mu_{xx}\right)^2 + 4\mu_2{}^2 \mu_{xy}{}^2 \tag{52}$$

$$B = 2k_0^2 \left[\left(\mu_2{}^2 - \mu_V \mu_{xx}\right)\left(\varepsilon_2 \mu_V - \varepsilon_1 \mu_2\right)\mu_2 \mu_{xx} - 2\varepsilon_2 \mu_2{}^3 \mu_{xy}{}^2\right] \tag{53}$$

$$D = -16k_0^4 \mu_{xy}{}^2 \mu_2{}^5 \mu_{xx} \left[\mu_2 \mu_{xx}\left(\varepsilon_1{}^2 + \varepsilon_2{}^2\right) - \varepsilon_1 \varepsilon_2 \left(\mu_2{}^2 + \mu_V \mu_{xx}\right)\right] \tag{54}$$

the asymptotic limits of $k_x$ can be calculated using the frequencies for which $A$ = 0. Furthermore, from the sign of *D*, it is possible to determine the regions over which a physical solution could exist but not all of the solutions of (51) are physically acceptable. In all of the figures shown, $R_1$ and $R_2$ define what will be termed the "the regions of existence". They are curves that define the boundaries between acceptable and unacceptable solutions based on the constraint that $\kappa$ in either material must be a real number. Importantly, they are the curves for which either $\kappa_1$ or $\kappa_2$ = 0. The labels 1 and 2 are used to define existence curves for YIG and LHM, respectively.

The data used for the dispersion curves are all given in terms of normalised parameters. For the YIG configuration, all of the frequencies are normalised with respect to $\omega_M$. In the cases where dispersion curves are shown for situations comprising only NPM, the frequencies are artificially normalized by a value of $\omega_M$ that will be used in the magnetised cases. This is so that all the dispersion curves have the same appearance for easier comparison. Similarly, the wave numbers have been normalised by a factor $k_0 = \omega_M / c$. The following set of results uses a saturation magnetisation, $4\pi M_0$ = 1750 G [42]. The analysis selected a relative permeability of the NPM of the order of -1 and this dictated that the normalised parameter $\omega_H / \omega_M$ = 1.43. Hence the value of the applied field is the order of 2500 G. The artificially normalised NPM parameters are $\omega_0 / \omega_M$ = 1.6423 and $\omega_{pe} / \omega_M$ = 2.0841. The principal results have been split to show the behaviour of each constituent layer together with the behaviour of the combined system.

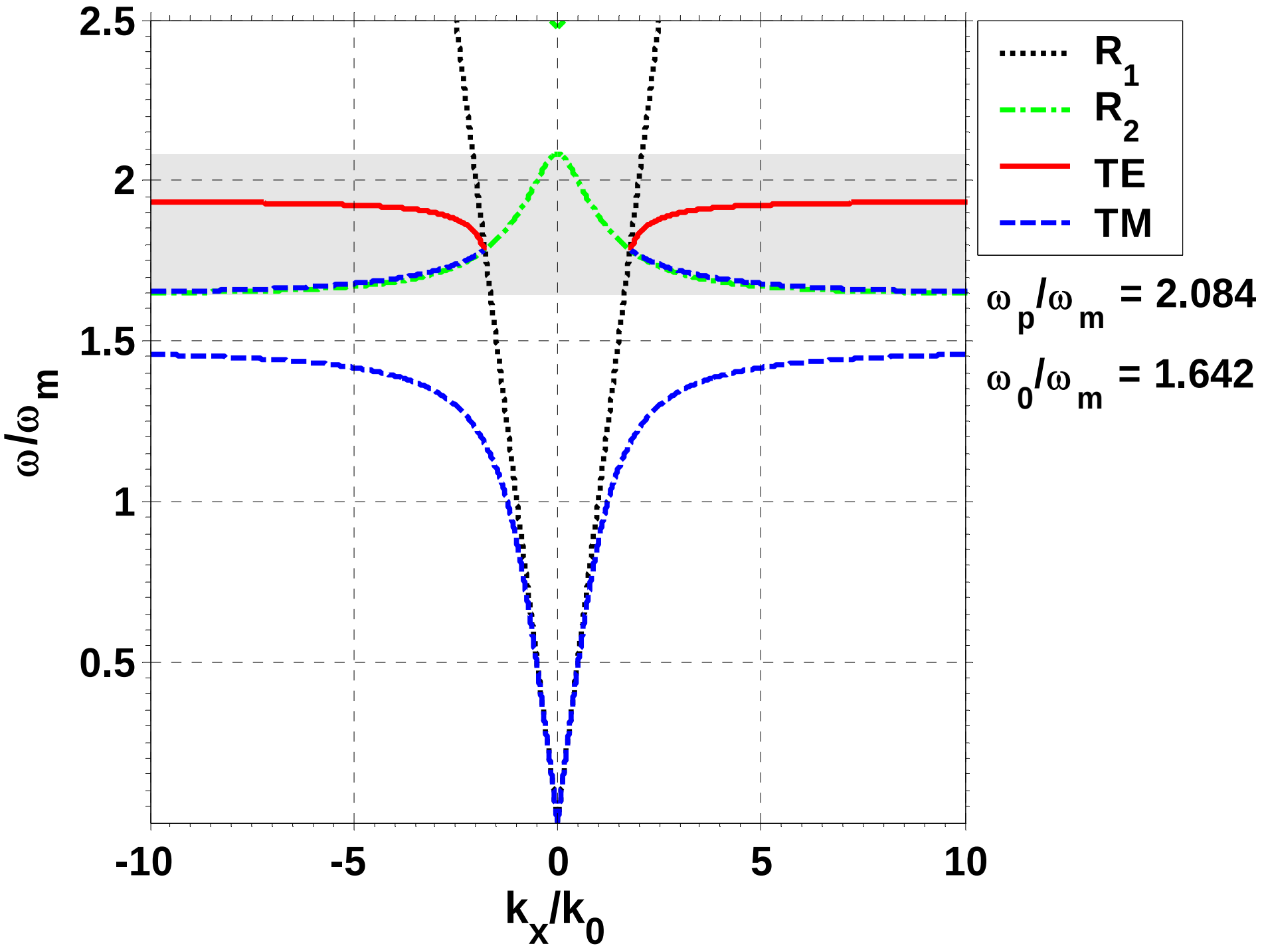

**Figure 4 Surface wave dispersion curve for an air-NPM interface, operating at GHz frequencies. Wave numbers are normalised by $k_0 = \omega_M / c$. Frequencies artificially normalized by a value $\omega_M$ that will be used for the magnetised cases to follow. The shading shows the electromagnetic transparent region where both the permittivity and the permeability are simultaneously negative.**

For these parameters the surface wave behaviour is reciprocal and the lower TM branch starts from zero. Nevertheless, in contrast to some previous results [13], the upper branch is now TE-polarised and *both* the upper TM and TE bands are within the transparent region.

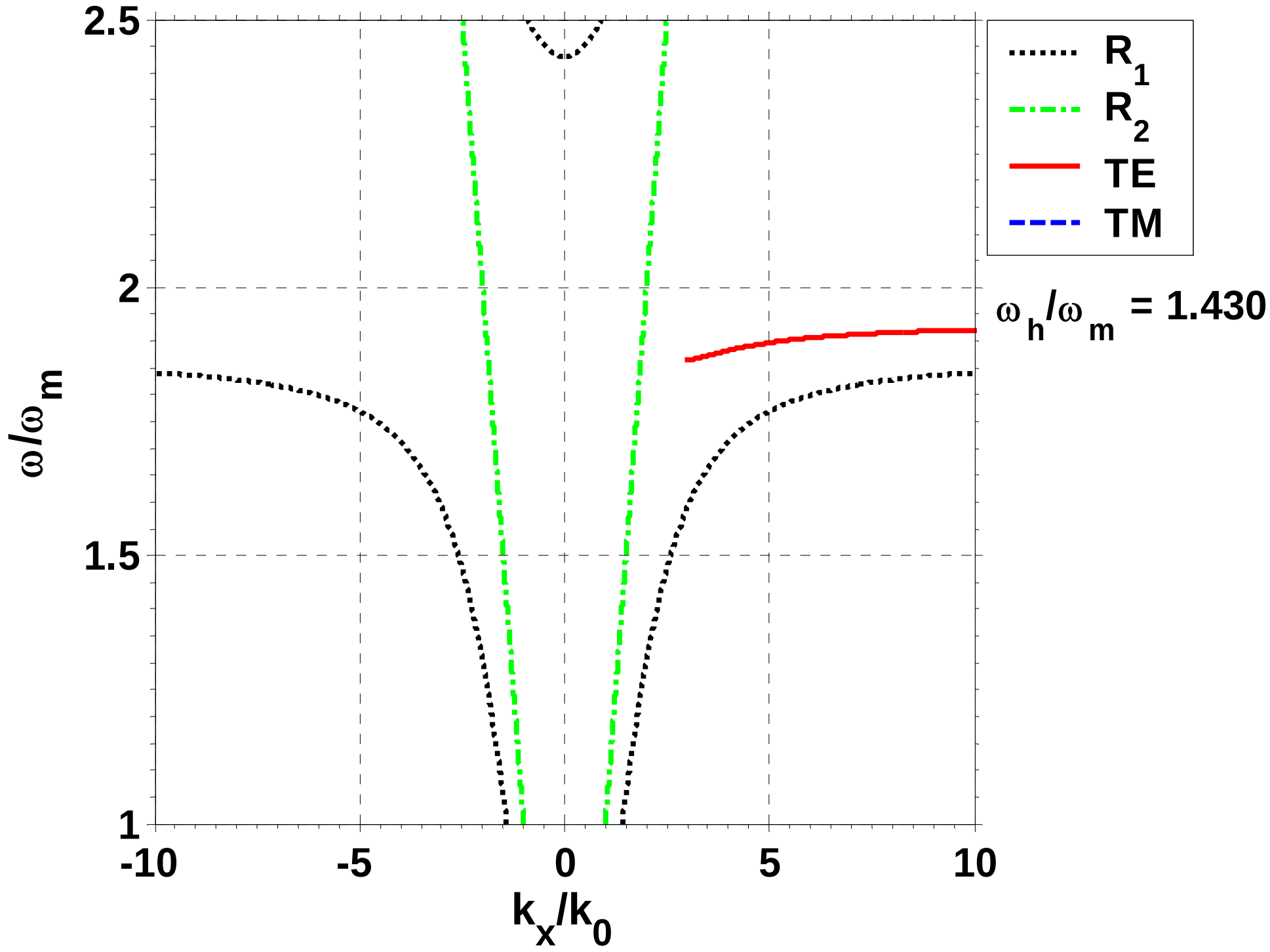

**Figure 5 Non-reciprocal TE-polarised surface wave dispersion curve for a YIG-air interface operating at GHz frequencies. Wave numbers and frequencies normalised as in Figure 4. Background relative permeability equal to unity.**

The only surface waves supported by a YIG-air interface are TE-polarised [24] and the dispersion curves in Figure 5 reveal the behaviour to be highly non-reciprocal. Note that the background permeability is unity. This is assumed because it is widely accepted for YIG but it should be noted that some very early work assumed that the background relative permeability was 1.25 and this step produces a limited branch in the negative wave number direction [24].

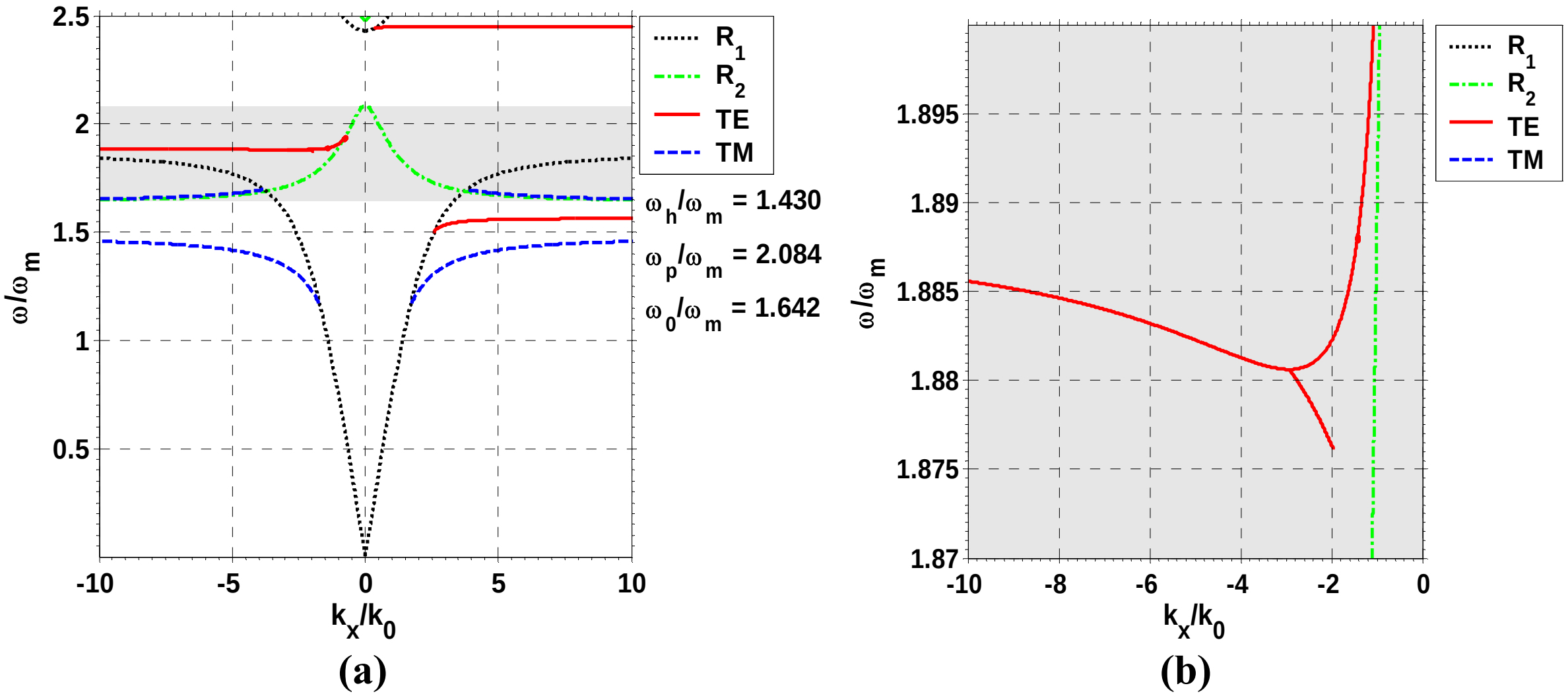

**Figure 6 Surface wave dispersion curves for (a) a YIG-NPM interface operating at GHz frequencies. Wave numbers and frequencies normalized as in previous figures. The shading shows the electromagnetic transparent region where both the permittivity and the permeability are simultaneously negative. (b) is an enlargement of the negative wave number TE-polarised branch.**

In this combined case, there exists both TE and TM modes but the TM modes are shifted from the NPM-air positions. This is understandable when it is remembered that $\mu$ does not explicitly arise in the dispersion equation for TM modes but it is nevertheless buried in the definition of $\kappa$. The TM branch now does not start from zero as it did in the NPM-air case. This is because $R_1$ and $R_2$ are now *both* functions of frequency and the behaviour of $R_1$ has acted to cut out the lower frequency TM solutions. It should be noted that the direction of the surface waves for the TM branches is unchanged from NPM-air.

For the TE modes, the presence of magnetised YIG has forced the NPM to acquire non-reciprocity but the latter is not as extreme as it is for the air-YIG interface because a negative wave number branch now exists. The enlarged image of the negative wave number TE branch shows that over a small frequency range the branch actually has *two* possible wave numbers for a given frequency. It is interesting that there is a switch from *positive* to *negative* $\frac{\partial \omega}{\partial k_x}$. This means that, within a given frequency range, it is possible to generate *simultaneously* a forward and backward surface wave.

## 4.2. Semi-infinite Negative Phase Medium bounded by a magneto-plasma

In this case, the dispersion equation (12) can be expanded using the following tensor elements

$$\varepsilon_{\mathrm{xx}} = \varepsilon_{\mathrm{yy}} = \varepsilon_{\infty}\left(1 - \frac{\omega_{\mathrm{p}}^{2}}{\left(\omega^{2} - \omega_{\mathrm{c}}^{2}\right)}\right), \quad \varepsilon_{\mathrm{xy}} = \varepsilon_{\infty}\frac{\omega_{\mathrm{c}}\omega_{\mathrm{p}}^{2}}{\omega\left(\omega^{2} - \omega_{\mathrm{c}}^{2}\right)} \tag{55}$$

$$\omega_{p}^{2} = \frac{Ne^{2}}{\varepsilon_{0}\varepsilon_{\infty}m^{*}}, \quad \omega_{c} = \frac{eB_{0}}{m^{*}} \tag{56}$$

where $N$ is the electron density, $m^*$ is the effective mass $e$ is the electronic charge, $B_o$ is the applied magnetic field and $\varepsilon_\infty$ is the background relative permittivity. Typically, the order of magnitude of the latter is $\varepsilon_\infty \sim 15$ for InSb [21,25]. Also, the magnetic field used to verify the reflectivity from a slab of n-type InSb can range from 0 to 10T. Given [21] that for InSb the electron density is $N \sim 10^{22}$ m$^{-3}$ and the effective mass is $m^* \sim 0.015 m_0$, the plasma frequency is typically $\omega_p \sim 10^{13}$ s$^{-1}$ and $\frac{\omega_c}{\omega_p} \sim B_o$. Hence experimentally the normalised parameter $\frac{\omega_c}{\omega_p}$ may lie in this case in the range 0 to 9. However this is also the range used in a recent measurement on GaAs/AlGaAs heterojunctions [43]

Once again, a biquadratic equation for $k_x$ emerges in a similar fashion to the GHz YIG instance. The parameters here have also been normalised but this time with respect to $\omega_p$. The wave numbers are therefore normalised by $k_0 = \omega_p / c$.

Figure 7(a) shows NPM operating at THz frequencies, where both permittivity and permeability are given by the Drude model. The TE branch is the upper branch and both branches lie below the region of transparency designated by the shaded area. As with the GHz NPM case, the THz dispersion is reciprocal. To give a better comparison, a dielectric-NPM configuration has been used such that the permittivity of the dielectric equals the background permittivity possessed by InSb, although it will be difficult to find a material with such a high frequency-independent permittivity. Figure 7(b) shows that only TM-polarised surface waves can be supported by an InSb-air surface. In addition, Figure 7(b) shows the kind of non-reciprocity that is present for a finite applied magnetic field. In fact, the effect of the magnetic field on the InSb-air system is to create a split set of TM branches [25] such that each $k_x$ direction possesses one asymptotic solution and one that terminates at a finite value.

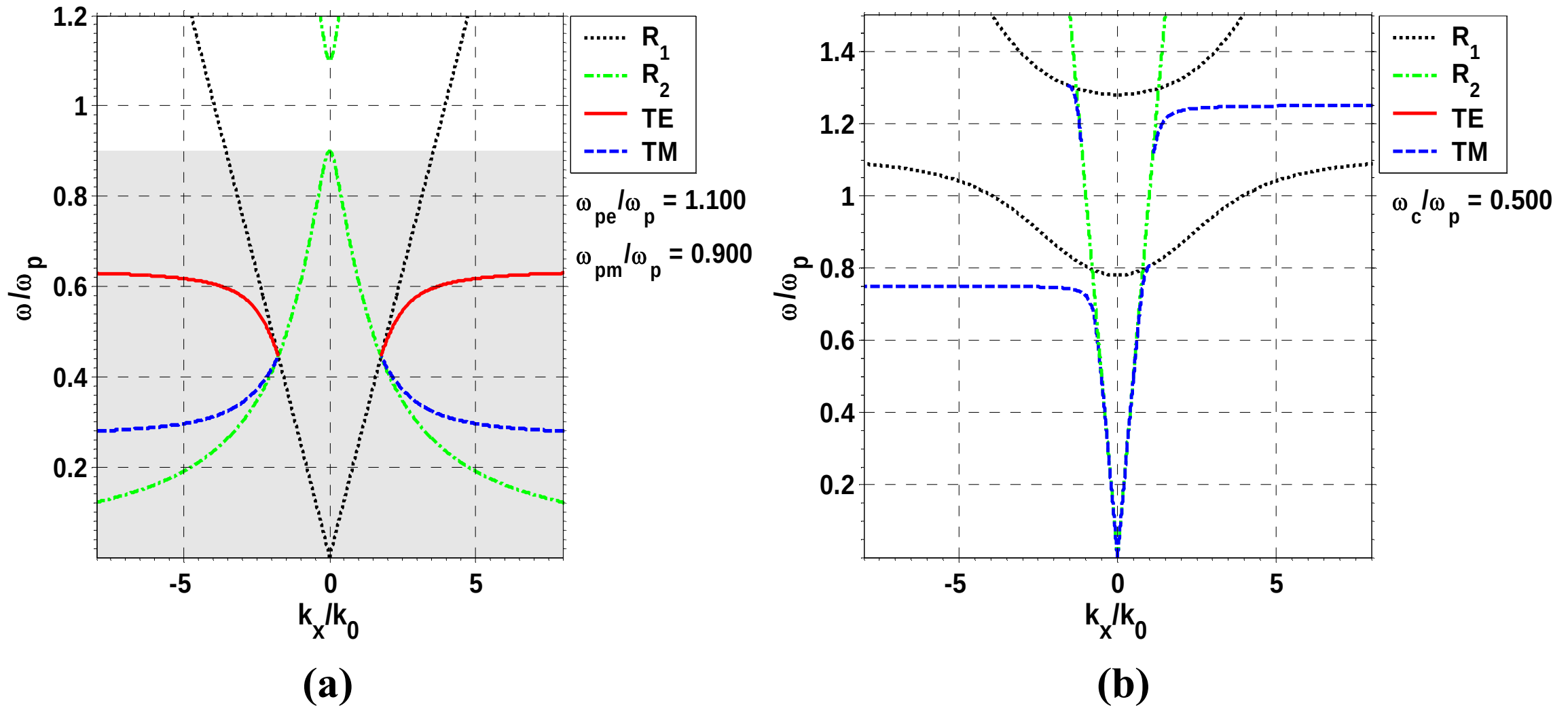

**Figure 7 Surface wave dispersion curves for (a) dielectric (ε = 15.68) -NPM and (b) InSb-air interfaces operating in an applied magnetic field at THz frequencies. Wave numbers normalised by $k_0 = \omega_p / c$ and frequencies by $\omega_p$. In (b) the strength of the magnetic field is given $\omega_c / \omega_p$. The shading indicates the transparency region.**

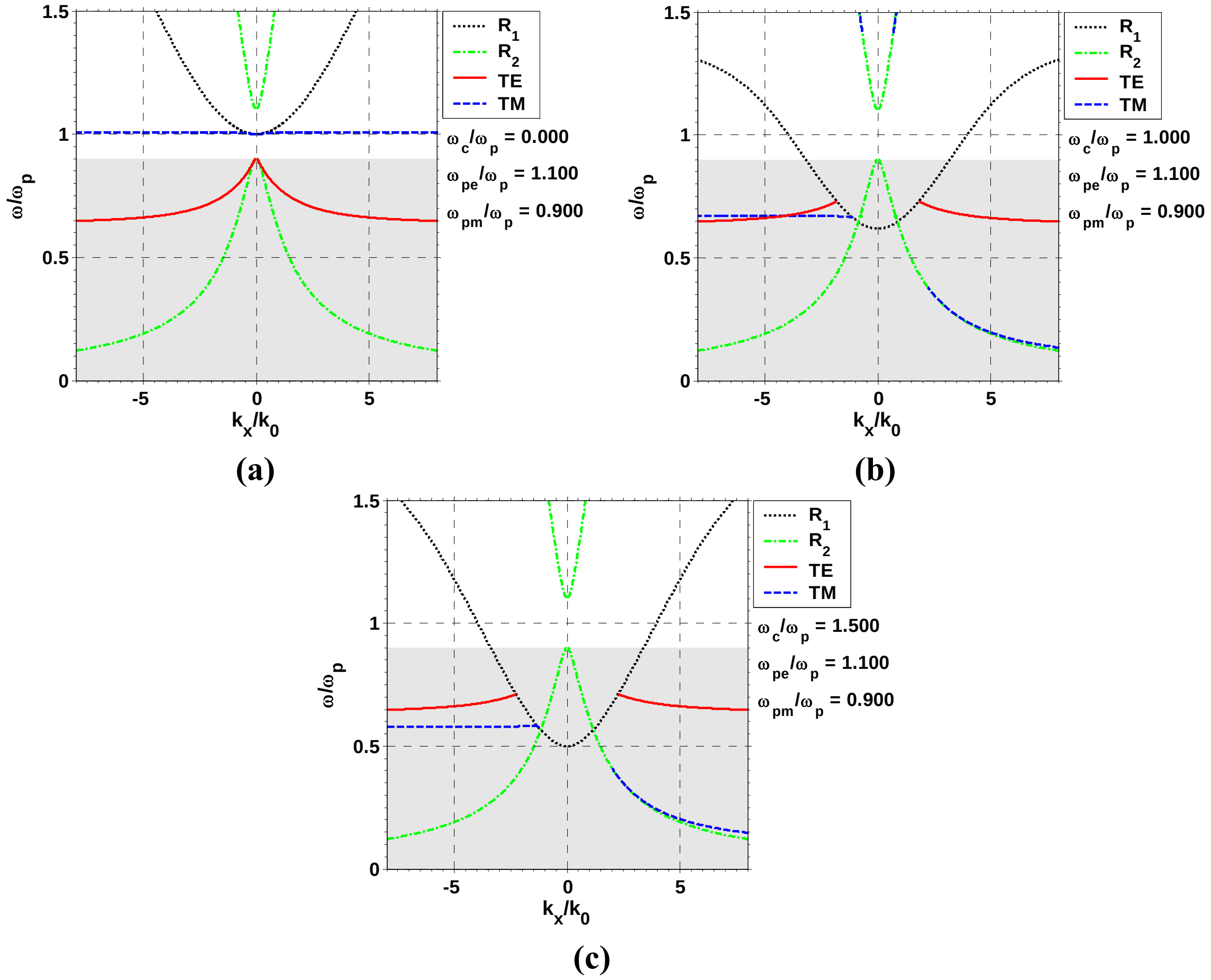

**Figure 8 Surface wave dispersion curves for InSb-NPM interface operating at THz frequencies. Wave numbers and frequencies normalised as in Figure 7. Figure (a) is with no applied field whereas (b) and (c) show the effect of increasing an applied field (increasing $\omega_c / \omega_p$). The shading indicates the transparency region.**

Figure 8 is a study of the InSb-NPM interface for various values of applied magnetic field. Figure 8(a) shows the combined effect of InSb and NPM at zero applied magnetic field. In this case, the dispersion offered by the NPM is being modified by a frequency dependent relative permittivity that is the same as that for a metal except that the background relative permittivity is now much greater than 1. It is clear that there is an obvious difference in the disposition of the TM- and TE-polarised branches between the artificial dielectric-NPM in Figure 7(a) and the zero magnetic field plasmonic permittivity deployed in Figure 8(a). The frequency dependence of the InSb relative permittivity, even at zero magnetic field, has pushed the TM branch out of the transparency region and has changed the property of the remaining TE branch.

Figures 8(b) and (c) show what happens for two values of the applied magnetic field, where the magnitude of the applied field can be deduced from the parameter $\omega_c/\omega_p$. First of all, non-reciprocity has appeared and a TM branch has been re-introduced into the transparency region. However, a striking new effect arises for the $-k_x$ branch where the TM mode is seen to sweep through and past the TE branch, as the applied magnetic field is increased.

It is remarkable that with the parameters used in Figure 8(a), the dispersion curves are allowed to start on the $k_x/k_0$ = 0 axis, since neither $R_1$ or $R_2$ are simple light lines. Figure 9 depicts the behaviour of the branches around the axis for the case of $\omega_c/\omega_p$ = 0. At such a point, the phase velocity, $v_p$, will be infinite however the figure clearly shows that the derivative at this point, and hence the group velocity is zero. This satisfies the well known condition $v_p v_g = const.$

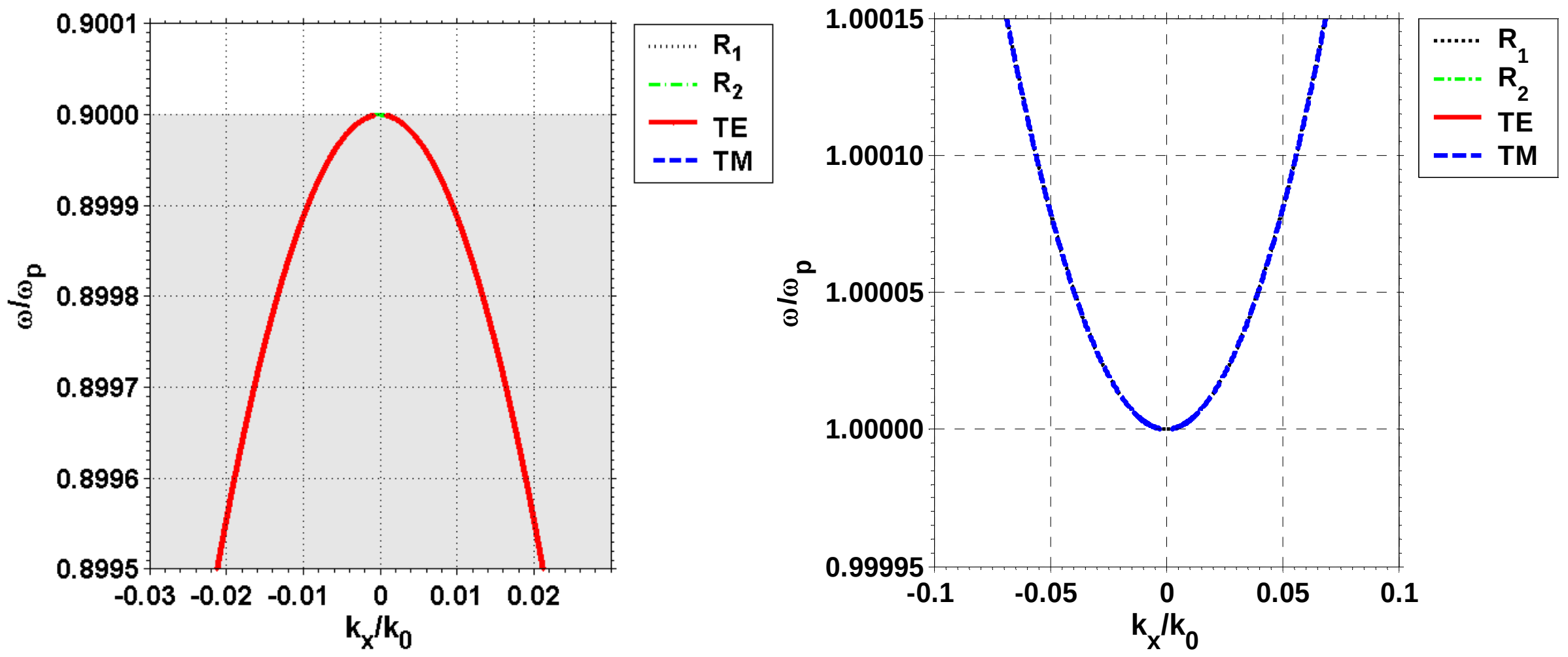

**(a)** **(b)**

**Figure 9 Magnified views of the surface wave dispersion near the wave number origin curve for InSb-NPM interface operating at THz frequencies with no applied magnetic field. Wave numbers normalised by $k_0 = \omega_p/c$, frequencies by $\omega_p$. Close to the point $k_x/k_0$ = 0 it can be seen that the curve does not form a cusp but instead goes to a point of zero gradient.**

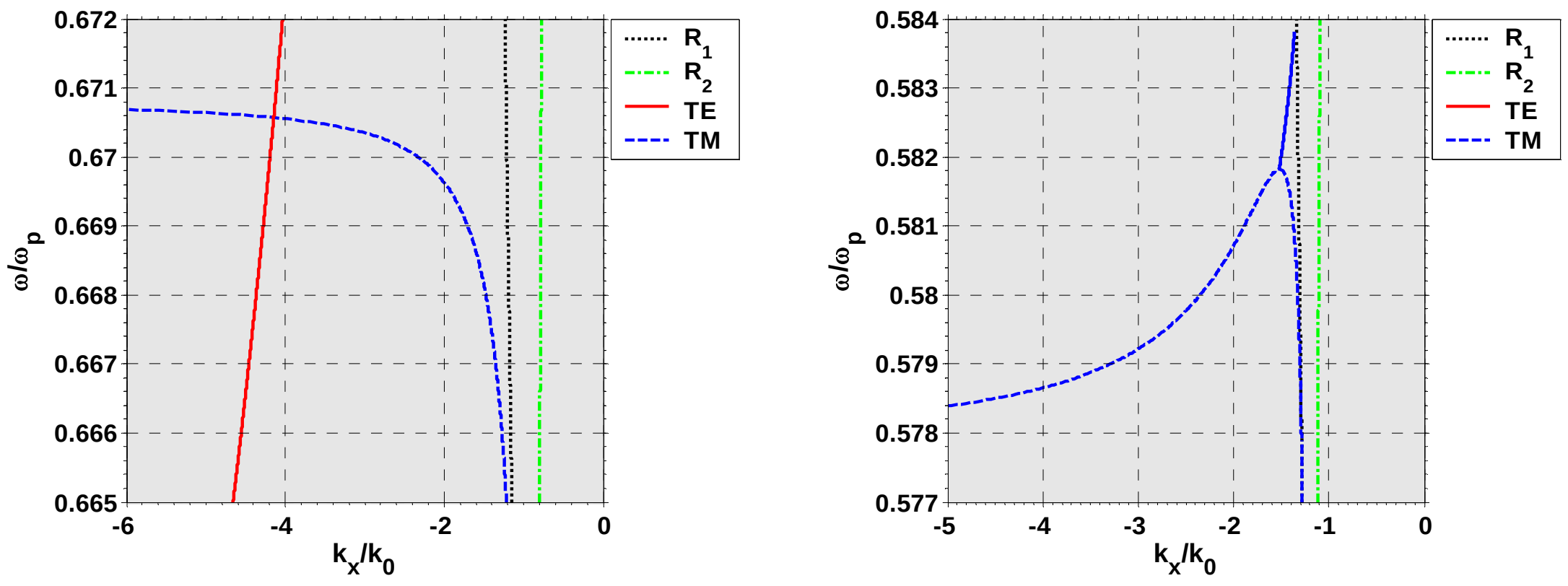

**(a)** **(b)**

**Figure 10 Magnified views of the surface wave dispersion curve for InSB-NPM interface operating at THz frequencies. Wave numbers normalised by $k_0 = \omega_p/c$, frequencies by $\omega_p$. Figures show the effect of increasing the applied field (effectively increasing $\omega_c$). Images are for (a) $\omega_c/\omega_p$ = 1.0 and (b) $\omega_c/\omega_p$ = 1.5 respectively.**

Figure 10 investigates what happens for two values of $\omega_c/\omega_p$. The TE and TM $-k_x$ branches can have solutions at the same frequency. This allows for the simultaneous generation of TE and TM surface modes and this is not possible using NPM alone. This would give rise to a very complex interaction between an unpolarised beam and this kind of interface. Above a certain applied field, the same $-k_x$ TM branch adopts a similar two-valued character that was as seen above in the YIG case. If the surface is interrogated with a pulse that has a broad enough frequency spectrum such that it cover the region occupied by the TE and TM branches then it should be possible to extract two separate frequency components, with opposing polarisations, from an unpolarised beam. This type of application will be subject of a further investigation.

### *4.3. Reflection simulations and Goos-Hänchen shifts*

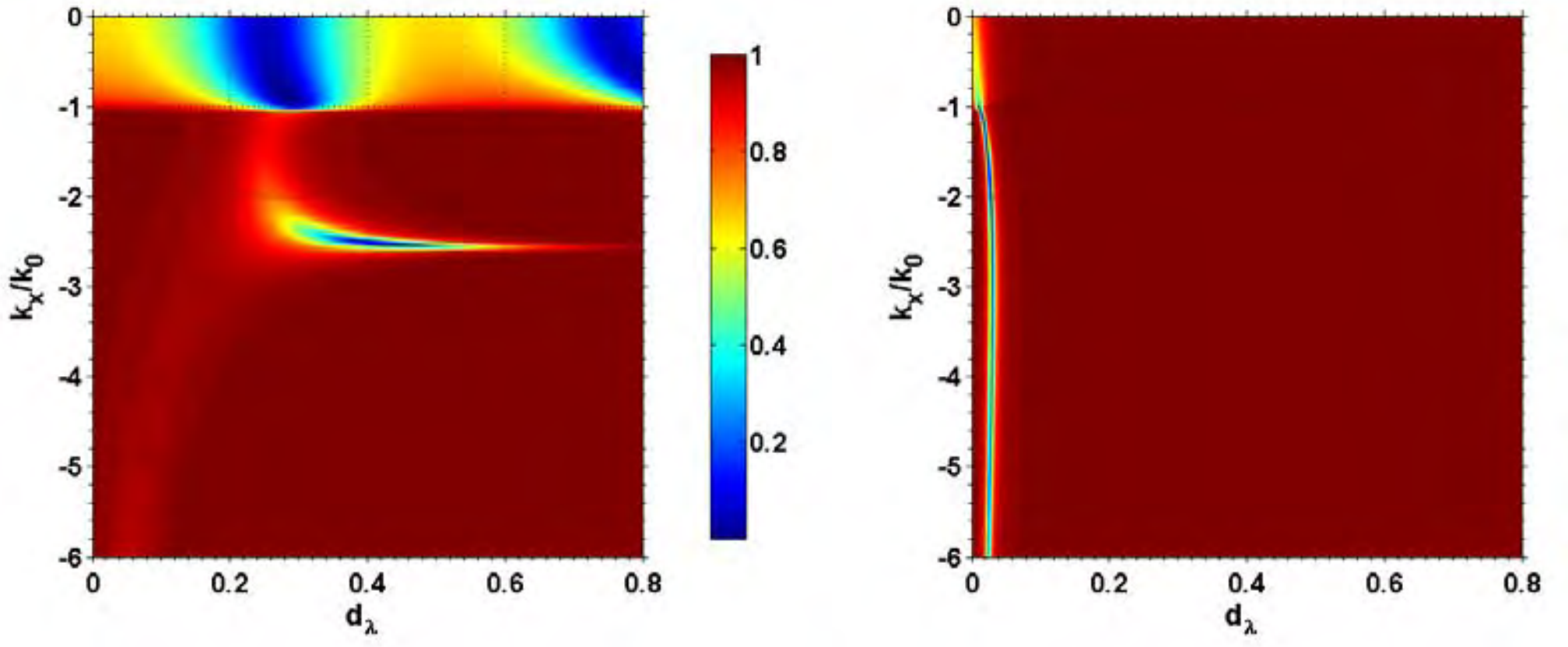

**(a)** **(b)**

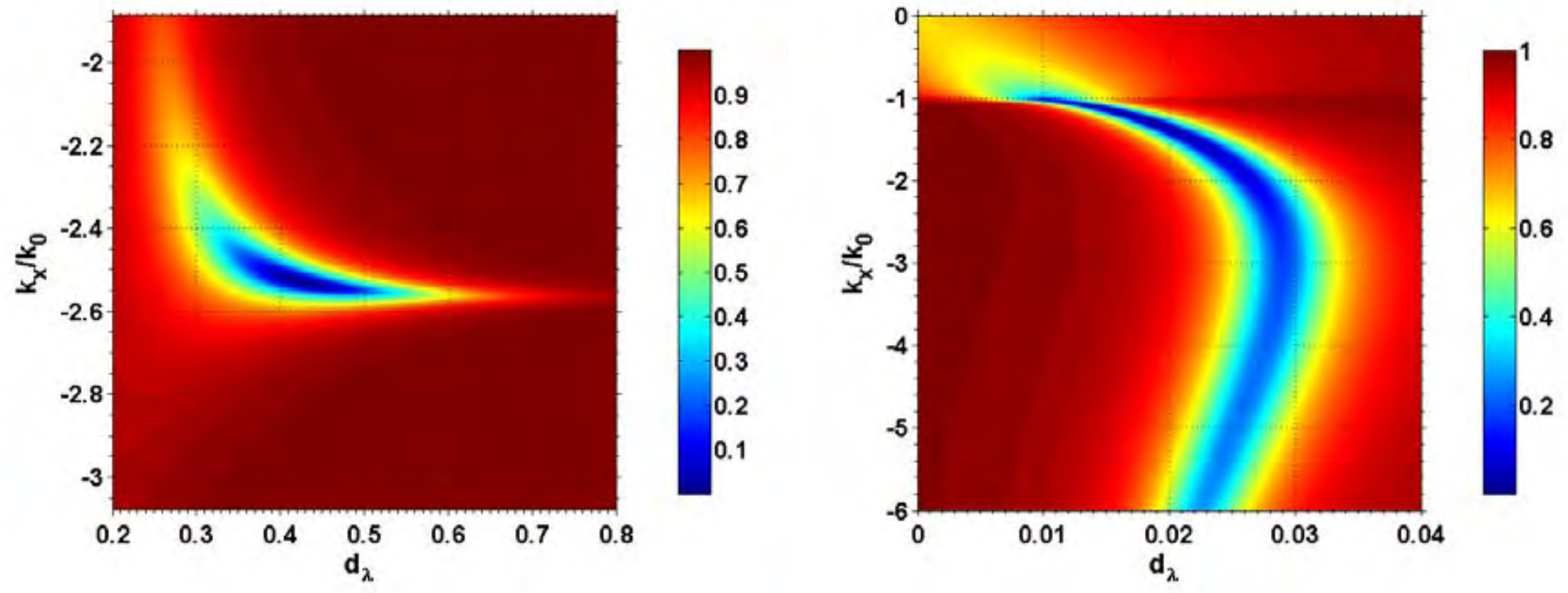

**(c)** **(d)**

**Figure 11 Calculated ATR reflected intensities using TE-polarised plane waves and for a prism-YIG-NPM configuration. Wave numbers are normalised by $k_0 = \omega_M / c$. YIG thickness is given in terms of $d_\lambda = d/\lambda_0$ where $\lambda_0$ is the free-space wavelength of the incident wave. (a) zero magnetic field. (b) finite magnetic field: $\omega_H / \omega_M$ = 1.43. $\omega / \omega_M$ = 1.882, $\omega_p / \omega_M$ = 2.084, $\omega_0 / \omega_M$ = 1.642, $\Gamma / \omega_M$ = 0.001 for all the results displayed. (c) and (d) are magnified versions of (a) and (b)**

Figure 11 displays in two dimensions the cross-section through the reflected intensity of a TE-polarised plane wave incident upon a standard ATR structure. The latter has been discussed earlier and the data shows that the reflectivity should reveal the presence of surface polariton resonances provided that some losses are included in the system. Accordingly, a loss parameter $\Gamma$ is introduced with the expectation that a resonance will be observed beyond the critical angle. In order to appreciate whether this and other normal ATR features are being generated it is necessary to make a comment about the broad general features of Figure 11. The plots are given in an interesting way that shows how to appreciate the dependence of the reflectivity upon the angle of incidence and the thickness of the YIG layer interposed between a prism (refractive index, $n$ = 3.87) and the type of NPM that operates at GHz frequency. This value of refractive index is rather high but it has been selected here so that in Figures 11(b) and (d), for a fixed value of $d_\lambda$ the two possible angles of incidence, first seen in Figure 6(b), can be readily revealed. In detail, the interpretation of Figure 11 is as follows. Take a fixed value of $d_\lambda$ and then progress upwards along the $k_x$-axis. A sequence of colour change will then be encountered that shows a progression from a reflectance of unity down through the minima associated with the resonances predicted by the dispersion curves.

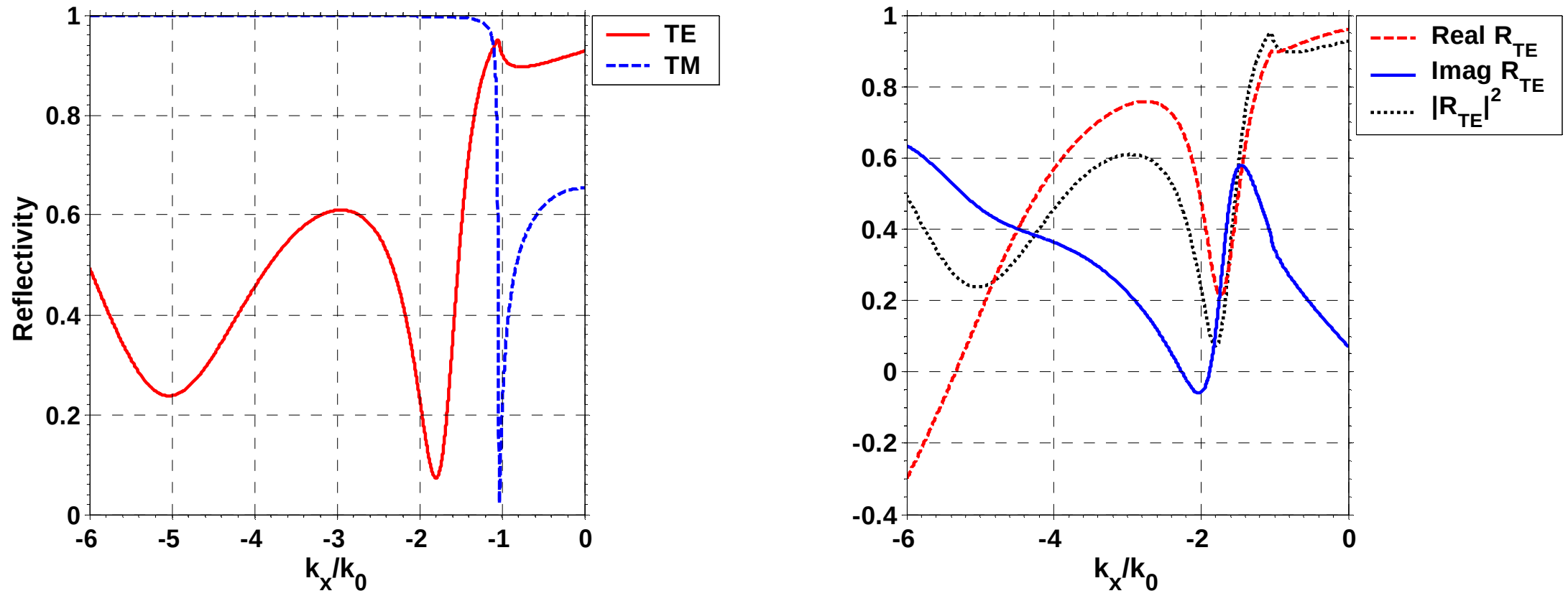

**(a)** **(b)**

**Figure 12 (a) Calculated TE and TM ATR reflected intensities for a prism-YIG-NPM configuration. (b) Real and imaginary parts of the TE reflected amplitudes and the TE reflected intensity. Wave vectors normalised by $k_0 = \omega_M / c$ YIG thickness given in terms of $d_\lambda = d / \lambda_0$ = 0.025 where $\lambda_0$ is the free space wavelength of the incident beam. $\omega_H / \omega_M$ = 1.43. $\omega / \omega_M$ = 1.882, $\omega_p / \omega_M$ = 2.084, $\omega_0 / \omega_M$ = 1.642, $\Gamma / \omega_M$ = 0.001 for all the results displayed.**

In addition to the plots seen in Figure 11, Figure 12(a) shows the variation of the reflectance as a function of $k_x$ for both the TE- and TM-polarisations. This information is a useful addition to the displays seen in Figure 11 because it shows the Brewster effect for the TM mode. In addition, Figure 12(b) gives a useful breakdown of how the real and imaginary parts of the TE reflected amplitude are changing.

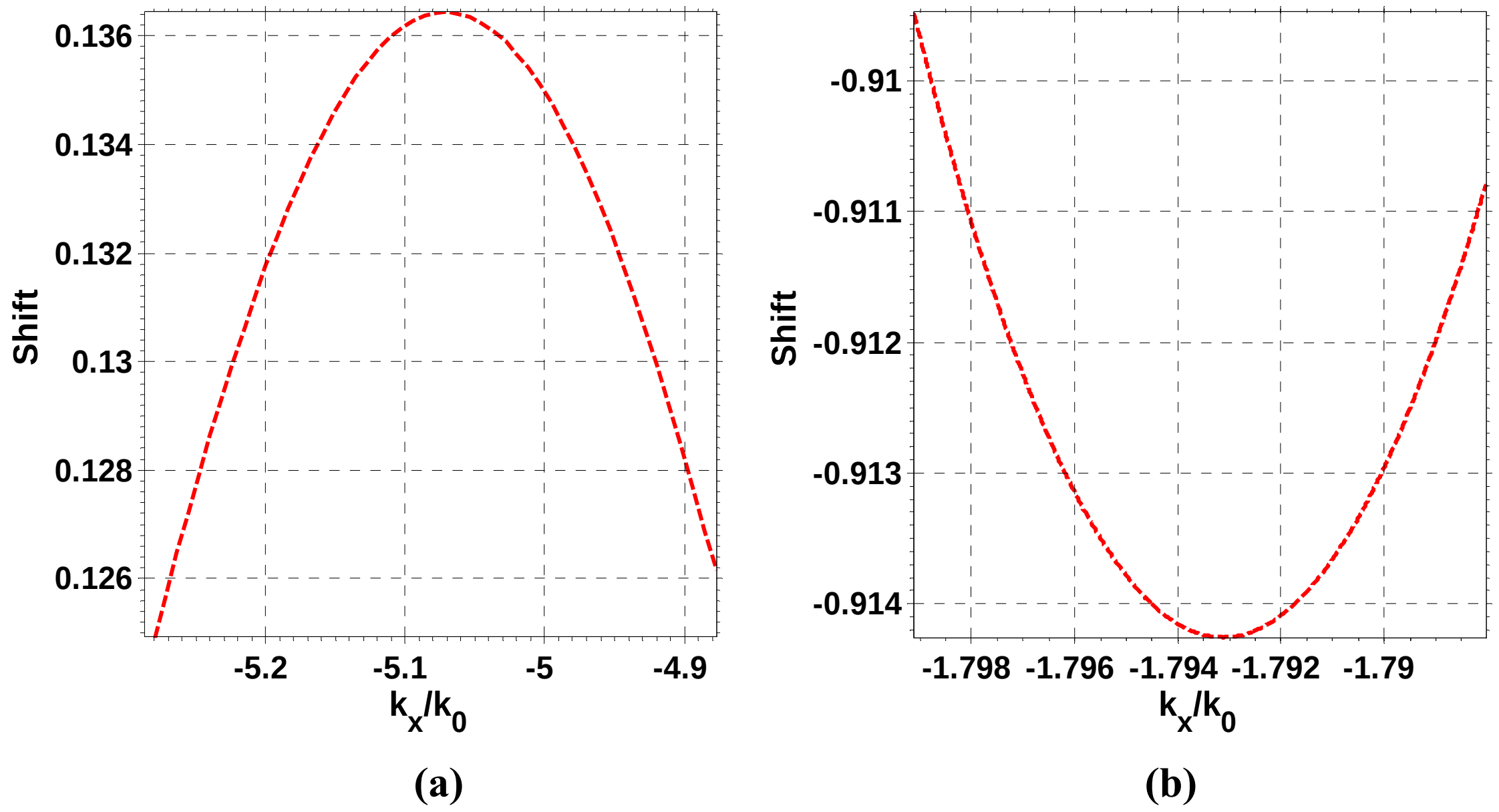

**Figure 13 Calculations of the factor $S_{GH}$, which is proportional to the Goos-Hänchen shift, for a TE beam incident on a prism-YIG-NPM configuration. Wave numbers normalised by $k_0 = \omega_M / c$. YIG thickness given in terms of $d_\lambda = d / \lambda_0$ = 0.025 where $\lambda_0$ is the free space wavelength of the incident beam. $\omega_H / \omega_M$ = 1.43. $\omega / \omega_M$ = 1.882, $\omega_p / \omega_M$ = 2.084, $\omega_0 / \omega_M$ = 1.642, $\Gamma / \omega_M$ = 0.001 for all the results displayed. (a) for the resonance at high wave number. (b) for the resonance at low wave number.**

Finally, Figure 13 shows an example of the main contribution to the Goos-Hänchen shifts. This is interesting because two distinct resonance possibilities are shown in Figure 12. As stated in the caption, Figure 13(a) is for a relatively large wave number. As seen in Figure 6(b), this is associated with a forward surface wave. Figure 13(b), for the lower wave number, is associated with a backward surface wave. Using the expression presented in (49), the behaviour of the Goos-Hänchen shifts can be seen to depend upon the magnitudes and derivatives of the real and imaginary parts of the reflected amplitude displayed in Figure 12. Upon performing the calculations it is immediately apparent that Figure 13(a) gives rise to a large positive shift and Figure 13(b) gives rise to a larger negative shift. These giant shifts are entirely associated with the presence of resonances and the existence of an NPM is not a necessary condition for the size of the shift. This is not surprising as pointed out earlier during the discussion of the Goos-Hänchen effect [40]. Another interesting point is that the shift can be positive, or negative, depending on the direction of the surface wave [44] but it is remarkable here that both of these shifts can be associated with the same dispersion curve.

## 5. Conclusions

This paper sets out to consider the role of gyrotropic media in the behaviour of surface waves and reflection systems that involve one of the new metamaterials that is called a negative phase velocity medium. The latter has often been referred to as left-handed medium. This name has arisen because the frequency region of interest involves a simultaneous use of a negative relative permittivity and a negative relative permeability. This property leads to negative refraction and it is important to know how this can be controlled by an external influence such as an applied magnetic field. The investigations reported here look in detail at how surface modes at the interface between a gyrotropic material respond in different frequency domains. The whole investigation uses what is called the Voigt configuration in which the surface modes propagate in a direction perpendicular to the field but nevertheless the applied magnetic field lies in the plane of the interface. A general theory of dispersion curves is given together with a comprehensive numerical assessment. The reflection of plane waves in a typical ATR configuration is also considered and the results are related to the resonances implied by the presence of surface excitations. This work is followed up by a consideration of what type of Goos-Hänchen shift is possible. It is shown that the magnitude of the shift is related to the resonances and that the sign of the shift is related to the details of some typical non-reciprocal dispersion curves. It is emphasised throughout that the latter are expected for this type of gyrotropic arrangement.